\documentclass[pra,twocolumn,titlepage,nofootinbib,amsmath,showpacs]{revtex4}%
\usepackage{graphicx}%
\newcommand{\Za}{Z\alpha}
\newcommand{\Zab}{(Z\alpha)}

\newcommand{\somevar}{\varepsilon}

\begin{document}
\title{Relativistic recoil effects to energy levels in a muonic atom: a Grotch-type calculation of the second-order vacuum-polarization contributions}
\author{Savely~G.~Karshenboim}
\email{savely.karshenboim@mpq.mpg.de}
\affiliation{Max-Planck-Institut f\"ur Quantenoptik, Garching,
85748, Germany} \affiliation{Pulkovo Observatory, St.Petersburg,
196140, Russia}
\author{Evgeny Yu. Korzinin}
\affiliation{D.~I. Mendeleev Institute for Metrology, St.Petersburg,
190005, Russia}
\author{Vladimir G. Ivanov} \affiliation{Pulkovo
Observatory, St.Petersburg, 196140, Russia}

\begin{abstract}
Adjusting a previously developed Grotch-type approach to a
perturbative calculation of the electronic vacuum-polarization
effects in muonic atoms, we find here the two-loop vacuum
polarization relativistic recoil correction of order
$\alpha^2(Z\alpha)^4m^2/M$ in light muonic atoms. The result is in
perfect agreement with the one previously obtained within the
Breit-type approach. We also discuss here simple approximations of
the irreducible part of the two-loop vacuum-polarization dispersion
density, which are applied to test our calculations and could be useful
for other evaluations with an uncertainty better than 1\%.
\end{abstract}
\maketitle

\section{Introduction}

High-precision tests of any advanced atomic calculations are
possible only for few-body systems and their accuracy is going down
dramatically when we increase the number of particles involved. The
highest accuracy has been achieved for two-body (hydrogen-like)
atomic systems. To study such systems not only are binding effects
and QED loops important but also recoil effects. While the
non-relativistic two-body problem is easily solved by introducing
the reduced mass, the relativistic recoil effects are more
complicated.

The problem of relativistic recoil effects was resolved for the pure
Coulomb two-body system a long time ago. The possible solutions
included the Breit equation (see, e.g., \cite{{breit1}}) and its
expansions as well as the Grotch equation \cite{Gro67}. For
non-Coulomb systems only the Breit-type approach has successfully
been used to date.

The purpose of this paper is to develop a method suitable for a
calculation of a certain class of corrections of order
$\alpha^2(Z\alpha)^4m^2/M$. The approach is applicable to medium-$Z$
muonic and antiprotonic atoms, i.e. to the atoms whose
characteristic atomic momentum $Z\alpha m$ is comparable to the
electron mass $m_e$. In such atoms the recoil effects are more
important than in ordinary (electronic) atoms. Meanwhile the
electronic vacuum polarization (eVP) effects are also enhanced. Thus
a calculation of relativistic recoil eVP corrections is important.
Here we calculate such corrections to the energy levels in the
second order of eVP effects, which are of order
$\alpha^2(Z\alpha)^4m^2/M$.

This papers is the third paper of the series \cite{I,II} devoted to a general
approach to calculate relativistic recoil effects and its
applications. In these papers, as explained in the
first paper of the series \cite{I}, we develop an approach which can be
applied for a certain class of potentials (or rather to a certain
class of corrections to the interaction between the atomic
particles). While the expressions are valid for a certain range of
atoms for arbitrary states, the practical importance of the
corrections depends on the value of the nuclear charge, the atomic
weight of the nucleus, the mass of the orbiting particle (which is
indeed different for muons and antiprotons) and the transition
between what levels are studied and with what accuracy.

At this stage we are interested in deriving the method and its
verification, rather than in its application to any particular
transition of practical interest. Below we derive the general
equations that take into account second-order-eVP relativistic recoil
effects. For the verification of the method we choose to
calculate the corrections which are known from a calculation
with an alternative (Breit-type) technique in our previous paper
\cite{a2Za4}.

Generalizing the method, developed by Grotch and Yennie \cite{Gro67} to
evaluate of the relativistic recoil effect in pure
Coulomb systems, to effects of eVP
in muonic atoms, in \cite{I,II} we derived the
general expression
\begin{eqnarray}\label{eq:1:exa}
E &=&m + m_R (f_{CN}({Z\alpha},Z\alpha m_R/m_e)-1)\nonumber\\
&-&\frac{m_R^2}{2M}\left(f_{CN}({Z\alpha},Z\alpha m_R/m_e)-1\right)^2\nonumber\\
&-&\frac{m_R^2}{2M}\frac{\partial}{\partial
\ln\kappa}\left(f_{CN}({Z\alpha},\kappa)-1\right)^2\Bigg\vert_{\kappa=Z\alpha m_R/m_e}\nonumber\\
&-&\langle \psi_{C} \vert  \left( \frac{V^2}{2M} + \frac{1}{4M}
[V,[\mathbf{p}^2,W]]
  \right) \vert \psi_{C} \rangle\,,
\end{eqnarray}
which is valid for an arbitrary perturbed potential
\[
V=V_C + V_N \,,
\]
where $V_C$ is the Coulomb potential and in a certain sense $V_N$ is
smaller than $V_C$, i.e. $V_N\sim \somevar V_C$, $\somevar\ll 1$. It
is important that $V_N$ is a kind of nonrelativistic potential
in the sense that its leading nonrelativistic contribution to the
energy is of order $\somevar(Z\alpha)^2m$, while the first
relativistic correction appears in order $\somevar(Z\alpha)^4m$.

Here, $W$ is a specific auxiliary potential, $\psi_{C}$ is the wave
function of the Dirac-Coulomb problem with the reduced mass, and
\begin{eqnarray}
f_{CN}({Z\alpha},Z\alpha m_R/m_e)&=&f_{C}({Z\alpha})\nonumber\\
&+&\Delta f_{CN}({Z\alpha},Z\alpha m_R/m_e)\;, \label{fcn}
\end{eqnarray}
is the exact dimensionless energy for the Dirac equation with the
reduced mass and potential $V$, and we separate the corrections to
it, $\Delta f_{CN}$, induced by $V_N$.

Expression (\ref{eq:1:exa}) is valid for non-recoil terms
exactly in $Z\alpha$, for the nonrelativistic problem exactly in
$m/M$ and for the leading relativistic recoil correction
$(Z\alpha)^4m^2/M$. It may be applied to an arbitrary order in
$\somevar$. (In principle, it may also be applied for an
$\somevar$ that is not small, if the appropriate wave functions and energy are found
numerically.)

\begin{figure}
\begin{center}
\includegraphics[width=0.12\textwidth]{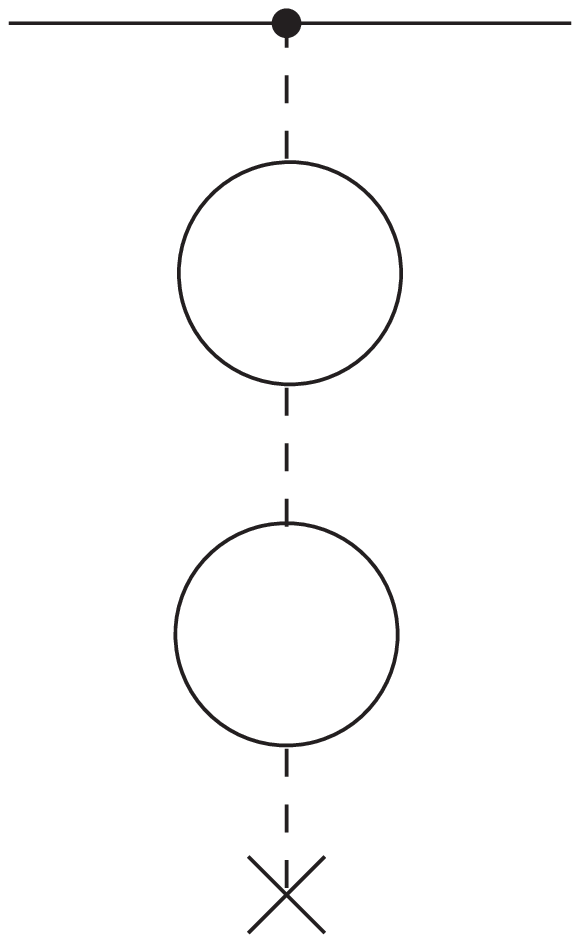}
\caption{Diagram of the reducible part of the K\"allen-Sabry
potential $V_{11}$.} \label{fig:2loopsa}
\end{center}
\end{figure}

\begin{figure}
\begin{center}
\includegraphics[width=0.25\textwidth]{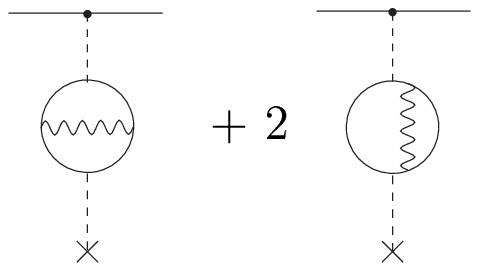}
\caption{Diagrams of the irreducible part of the K\"allen-Sabry
potential $V_{2}$.} \label{fig:2loopsb}
\end{center}
\end{figure}

In \cite{II} we describe a method to calculate the recoil correction
to the energy of order $\alpha/\pi\Zab^4m^2/M$, and here we aim to
obtain the recoil correction of the next order in $\alpha$.  For
that we consider a potential of the form
\begin{equation}\label{eq:vvv}
  V_N = V_U + V_{11} + V_2
  \,,
\end{equation}
where
\[
V_U=-Z\alpha \int_0^1 dv \, \rho_1(v) \frac{e^{-\lambda r}}{r}
\]
is the Uehling potential,
\[
V_{11}=-Z\alpha \int_0^1 dv \, \rho_{11}(v) \frac{e^{-\lambda r}}{r}
\]
corresponds to the reducible two-loop eVP potential (see
Fig.~\ref{fig:2loopsa}) and
\[
V_{2}=-Z\alpha \int_0^1 dv \, \rho_2(v) \frac{e^{-\lambda r}}{r}
\]
is for  its irreducible part (see Fig.~\ref{fig:2loopsb}).  The
dispersion parameter is
\[
\lambda^2=\frac{4m_e^2}{1-v^2}\;,
\]
and the eVP dispersion density functions are defined as
\cite{Schwinger,Kallen-Sabry,ro2hfs,muhfs}
\begin{eqnarray}
 \rho_{1}(v)&=&\left(\frac{\alpha}{\pi}\right)\frac{v^2(1-v^2/3)}{1-v^2}\,,
\end{eqnarray}
\begin{eqnarray}
 \rho_{11}(v)&=& - \frac{1}{9}\left(\frac{\alpha}{\pi}\right)^2\,\frac{v^2(1-v^2/3)}{1-v^2}
 \nonumber\\
 &\times&
 \left\{16-6v^2+3v(3-v^2)\ln\left(\frac{1-v}{1+v}\right)\right\}\;,
\end{eqnarray}
\begin{eqnarray}
  \rho_{2}(v)&=&\frac{2}{3}\left(\frac{\alpha}{\pi}\right)^2\frac{v}{1-v^2}\times\biggl\{(3-v^2)(1+v^2)\nonumber\\
   &&\left[{\rm Li}_2\left(-\frac{1-v}{1+v}\right)+2{\rm Li}_2\left(\frac{1-v}{1+v}\right)\right.\nonumber\\
  &+&\left. \ln\left(\frac{1+v}{1-v}\right)\left(\frac{3}{2}\ln\left(\frac{1+v}{2}\right)  -\ln\left(v\right)\right)\right]\nonumber\\
  &+&\left(\frac{11}{16}(3-v^2)(1+v^2)+\frac{1}{4}v^4\right)\ln\left(\frac{1+v}{1-v}\right)\nonumber\\
  &+&\frac{3}{2}v(3-v^2)\ln\left(\frac{1-v^2}{4}\right)-2v(3-v^2)\ln(v)\nonumber\\
  &+&\frac{3}{8}v(5-3v^2)\biggr\}\,,
\end{eqnarray}
where ${\rm Li}_2(z)$ is the Euler dilogarithm \cite{Lewin}.

\begin{figure}
\begin{center}
\includegraphics[width=0.18\textwidth]{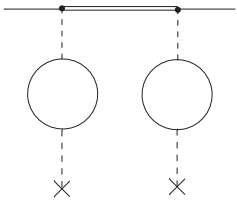}
\caption{The second-order Uehling contributions to the energy.} \label{fig:ugu}
\end{center}
\end{figure}

It is important that Eq.~(\ref{eq:1:exa}) includes only the first
derivative with respect to $\kappa$. That is because a shift in the effective
mass is at most ${\cal O}((Z\alpha)^2(m/M)m)$ and terms quadratic in
the shift are at most of the order $\somevar(Z\alpha)^6(m/M)^2m$. To
evaluate the derivative we apply the identity
\[
\frac{\partial \left(f_{CN}({Z\alpha},\kappa)-1\right)^2}{\partial
\ln\kappa}=-\frac{\partial \left(f_{CN}({Z\alpha},Z\alpha
m_R/m_e)-1\right)^2}{\partial \ln m_e}\;,
\]
which allows us to avoid differentiating the $\alpha^2$ term in
$f_{CN}({Z\alpha},Z\alpha m_R/m_e)$, found by means of numerical
computation, and instead allows to calculate numerically an integral, which
contains a derivative of the potential over the parameter $m_e$.

The result for the Dirac equation with the reduced mass and the
potential defined in Eq. (\ref{eq:vvv}) can be in principle
obtained by many means. Recently such a result for low-lying states
in light muonic atoms was found by applying a
nonrelativistic perturbation theory in \cite{a2Za4}.

\section{eVP relativistic recoil corrections to the second order of $\alpha$}

Let us expand Eq.~(\ref{fcn}) in terms of $\somevar$:
\begin{eqnarray}
f_{CN}({Z\alpha},Z\alpha m_R/m_e)&=&f_{C}({Z\alpha})
+ f_{N}({Z\alpha},Z\alpha m_R/m_e)\nonumber\\
&&+f_{NN}({Z\alpha},Z\alpha m_R/m_e)\;,
\end{eqnarray}
where $f_N$ is linear in $\somevar$ and $f_{NN}$ is quadratic.

It is convenient to consider different parts of
the perturbing potential independently,
setting appropriate $\somevar$ in different cases.

\subsection{The first-order contribution of $V_{11}$ and $V_2$}

To take into account the contributions of $V_{11}$
and $V_2$, we can set $\somevar=(\alpha/\pi)^2$ and take the first
order of the perturbation theory (in $\somevar$), which has been already studied in Ref. \cite{II}%
\footnote{To simplify notation we drop the arguments in terms of
Eq.~(\ref{eq:1:exa}).}
\begin{equation}
E_{Nq}^{(1)} = m_R f_{Nq} + \Delta
E_{q}^{(1)} \,,
\end{equation}
where the index $q$ corresponds to either $V_{11}$ or $V_{2}$,
\begin{eqnarray}\label{eq:1:all}
\Delta E_{q}^{(1)}&=&-\frac{m_R^2}{M}\Bigl(f_{C}-1\Bigr)f_{Nq}\nonumber\\
&-&\frac{m^2_R}{M} (f_C-1) \frac{\partial}{\partial
\ln\kappa}f_{Nq}\nonumber\\
&-&\langle \psi_C |  \Bigl(
 \frac{V_q V_C}{M}
    + \frac{1}{4M} [V_C, [\mathbf{p}^2, W_q] ]
\nonumber\\
 && ~~ + \frac{1}{4M} [V_q, [\mathbf{p}^2, W_C] ]\Bigr) | \psi_C
 \rangle\,,
\end{eqnarray}
$|\psi_C\rangle$ is the Coulomb wave function and it is sufficient to consider it in the nonrelativistic (NR) approximation (cf. \cite{II}).
The related auxiliary potential takes the form (cf. (10) of \cite{II})
\begin{eqnarray}\label{wu1}
    W_q({\mathbf k})&=& 8\pi\Zab \int_0^1
  dv \,  \frac{\rho_q(v)}{({\mathbf k}^2+\lambda^2)^2}
  \,,\nonumber\\
  W_q(r) &=& {\Zab} \int_0^1
  dv \, \rho_q(v) \frac{e^{-\lambda r}}{\lambda}
  \,.
\end{eqnarray}

Proceeding in the same way as in \cite{II} (cf. Eq.~(32) there), we
obtain for these contributions (cf. \cite{rel1,rel2})
\begin{eqnarray}\label{deu_nr}
  \Delta E_{Nq}^{\rm (NR)}
  &=&
  \frac{\Zab^4}{n^3} \frac{m_R^2}{M}
  \sum_{i,k=0}^{n-l-1}
  B_{ik}^{\rm (NR)}\nonumber\\&\times& 
  \Biggl[
    -\frac{1}{2n} K^{(q)}_{2,2l+i+k+2}(\kappa_n)\nonumber\\
    &&\phantom{9}
    -\frac{2l+i+k+2}{2n\kappa_n} K^{(q)}_{3,2l+i+k+3}(\kappa_n)\nonumber\\
    &&\phantom{9}
    +\frac{1}{\kappa_n} K^{(q)}_{3,2l+i+k+2}(\kappa_n)
  \Biggr]
  \,,
\end{eqnarray}
where
\begin{eqnarray}
  B_{ik}^{\rm (NR)}
  &=&
  \frac{(-1)^{i+k}(n-l-1)!}{i!(n-l-i-1)!k!(n-l-k-1)!} 
  \nonumber\\
  &\times&
  \frac{(n+l)!(2l+i+k+1)!}{(2l+i+1)!(2l+k+1)!}
\end{eqnarray}
and functions
\begin{eqnarray}
  K^{(q)}_{bc}(\kappa)
  &=&
  \int_{0}^{1}{dv}\, \frac{\rho_q(v)}{(1-v^2)^{b/2-1}}\,
  \left(\frac{\kappa\sqrt{1-v^2}}{1+\kappa\sqrt{1-v^2}}\right)^{c}
\end{eqnarray}
differ from the base integrals $K_{bc}(\kappa)$ \cite{rel1,rel2},
introduced earlier and expressed in terms of spectral functions.

For the low-lying states of interest ($n=1,2$) in light muonic atoms
the results are
\begin{eqnarray}
  \Delta E_{q}^{\rm (NR)} (1s) &=&
  \Zab^4\, \frac{m_R^2}{M}\,\nonumber\\
  &\times&  \frac{1}{\kappa} \biggl[-\frac{\kappa}{2}K^{(q)}_{22}(\kappa)+K^{(q)}_{32}(\kappa)-K^{(q)}_{33}(\kappa)  \biggr]
   \;,\nonumber\\
 \Delta E_{q}^{\rm (NR)} (2s) &=& \Zab^4\, \frac{m_R^2}{32M} \nonumber\\
  &\times& \Biggl\{-K^{(q)}_{24}(\kappa_2)+\frac4{\kappa_2}\Bigl[ K^{(q)}_{34}(\kappa_2)-K^{(q)}_{35}(\kappa_2) \Bigr] \nonumber\\
  &+&\frac{2}{\kappa_2^3}\Bigl[ \kappa_2 K^{(q)}_{44}(\kappa_2)+4K^{(q)}_{54}(\kappa_2) \nonumber\\
  &&-4K^{(q)}_{55}(\kappa_2)\Bigr]\Biggr\}
  \;,\nonumber\\
  \Delta E_{q}^{\rm (NR)} (2p) &=&  \Zab^4\,  \frac{m_R^2}{M}\,\nonumber\\
  &\times&  \frac{1}{\kappa_2}  \biggl[  -\frac{\kappa_2}{2}K^{(q)}_{2,4}(\kappa_2)+2 K^{(q)}_{3,4}(\kappa_2)\nonumber\\
  &&  -2 K^{(q)}_{3,5}(\kappa_2)  \biggr]
  \,.
\end{eqnarray}
The required integrals $K^{(11)}_{bc}$ and $K^{(2)}_{bc}$ should be
calculated numerically. The numerical results are considered in
Sec.~\ref{s:res}.

\subsection{The second-order contribution of $V_U$}

To deal with the second order contributions of $V_U$ (see, e.g,
Fig.~\ref{fig:ugu}) we should address terms in (\ref{eq:1:exa})
which are the second order in $\somevar=\alpha/\pi$.

For the case of general $V_N$, selecting
terms of the corresponding order $\somevar^2$ in Eq.~(\ref{eq:1:exa}), we arrive
at
\begin{equation}\label{eq:2ab}
  E^{(2)}_N
  =
  m_R f_{NN}
  +
  \Delta E^{(2a)}_N
  +
  \Delta E^{(2b)}_N
  \,,
\end{equation}
where
\begin{eqnarray}\label{fb2}
\Delta E^{(2a)}_N
=
-\frac{m_R^2}{M} \left(\vphantom{\frac00} (f_C-1)f_{NN}
+ \frac{\bigl(f_N\bigr)^2}{2}\right)
\nonumber\\
-\frac{m_R^2}{M}\left((f_C-1)\frac{\partial}{\partial
\ln\kappa}f_{NN} +f_N \frac{\partial}{\partial \ln\kappa}
f_N\right)\;.
\end{eqnarray}
Note that in the last expression we need only the leading
nonrelativistic contribution to various $f$, and in particular it
is sufficient to write
\begin{equation}
f_{NN} = \frac{\langle \psi_C \vert V_N G^\prime_C V_N\vert
\psi_C\rangle}{m_R}\;,
\end{equation}
where $G^\prime_C$ is the reduced Coulomb wave function for the
corresponding state.

The second term in (\ref{eq:2ab}), $\Delta E^{(2b)}$, corresponds to the
$W$ contribution of the matrix element in (\ref{eq:1:exa}), which in
the second order of $\somevar$ provides us with
\begin{eqnarray}
\Delta E^{(2b)}_N
&=&
-\langle \psi_C \vert  \left( \frac{V_{N}^2}{2M} + \frac{1}{4M}
 [V_{N},[\mathbf{p}^2,W_{N}]]
  \right) \vert \psi_C \rangle
\nonumber\\
&-&2\langle \psi_C \vert  \left( \frac{V_CV_{N}}{M}+\frac{1}{4M}
 [V_N,[\mathbf{p}^2,W_C]]
 \right.
\nonumber\\
 &+& \left.\frac{1}{4M}
 [V_C,[\mathbf{p}^2,W_N]]
  \right) \vert \psi_N \rangle
  \,,
\end{eqnarray}
where
\begin{equation}\label{psin}
|\psi_N\rangle=G^\prime_C V_N |\psi_C\rangle
\end{equation}
is the correction to the wave function induced by $V_N$.

To calculate the correction to the second order in the Uehling
potential we set $V_N=V_U$ in general
expressions~{(\ref{fb2})--(\ref{psin})}.

Equation (\ref{eq:2ab}) presents the complete result
for the relativistic terms of order $\alpha^2(Z\alpha)^4m$ and
$\alpha^2(Z\alpha)^4m^2/M$ in terms of the sum of the Dirac term
with the reduced mass $m_R f_{NN}$ and the recoil corrections $\Delta
E^{(2a)}_U$ and $\Delta E^{(2b)}_U$. Since the expressions for the
corrections deal with nonrelativistic wave functions and Green
functions, the result for the eVP relativistic recoil correction
depends on orbital momentum $l$ and does not depend on total muon
angular momentum $j$, which means that there is no corrections to
the fine splitting in this order behind the result of the Dirac
equation with the reduced mass.

\section{Results\label{s:res}}

The expressions presented above allow us to calculate the recoil
correction to the energy of order $(\alpha/\pi)^2\Zab^4 m_R^2/M$.
The contributions to the correction for the lowest states of the muonic hydrogen
\begin{equation}\label{eq:1122ab}
\Delta E^{\rm (rec-VP2)}= \Delta E^{(1)}_{11} +\Delta E^{(1)}_{2}
+\Delta E^{(2a)}_{U} +\Delta E^{(2b)}_{U}
\end{equation}
are listed for muonic hydrogen in Table~\ref{tab1}.

\begin{table}
\begin{center}
\begin{tabular}{|c|c|c|c|}
\hline
& $1s$ & $2s$ & $2p$\\
\hline
$\Delta E^{(1)}_{11}$ & $\phantom{-}0.505$  & $\phantom{-}0.0721$  & $\phantom{-}0.000133$ \\
$\Delta E^{(1)}_{2}$  & $\phantom{-}0.139$  & $\phantom{-}0.0576$  & $\phantom{-}0.002879$ \\
$\Delta E^{(2a)}_{U}$ & $-0.154$ & $-0.0024$ & $-0.000060$ \\
$\Delta E^{(2b)}_{U}$ & $\phantom{-}0.678$  & $\phantom{-}0.0685$  & $\phantom{-}0.000367$ \\
\hline
$\Delta E^{\rm (rec-VP2)}$ &$\phantom{-}1.168$ & $\phantom{-}0.1958$ & $\phantom{-}0.003319$ \\
\hline
\end{tabular}
\caption{The $\alpha^2$ eVP relativistic recoil corrections to
energies of the muonic hydrogen in units of $(\alpha/\pi)^2\Zab^4
m_R^2/M$. See Eq.~(\ref{eq:1122ab}) for notation.} \label{tab1}
\end{center}
\end{table}

The first order K\"allen-Sabry potential contributions are
calculated numerically in a rather straightforward way. To control the
calculation of the irreducible part we have also used  various approximate
representations for the dispersion function. They are discussed in
Appendix \ref{a:ks}.

The corrections of the second order in the Uehling potential were
computed using two different representations of the reduced
nonrelativistic Coulomb Green functions, which are summarized in
Appendix~\ref{a:g}. The results produced with two representations
are consistent.

The calculations were also performed for various isotopes of muonic
hydrogen and helium ion. The results are presented in
Table~\ref{tab2}.

\begin{table}
\begin{center}
\begin{tabular}{|c|c|c|c|}
\hline
& $1s$ & $2s$ & $2p$\\
\hline
H          & $1.168$ & $0.1958$ & $0.003319$ \\
D          & $1.192$ & $0.2016$ & $0.003551$\\
$^3$He$^+$ & $1.445$ & $0.2867$ & $0.006866$ \\
$^4$He$^+$ & $1.447$ & $0.2878$ & $0.006900$\\
\hline
\end{tabular}
\caption{The $\alpha^2$ relativistic recoil eVP corrections to
energies of the light muonic hydrogen-like atoms in units
$(\alpha/\pi)^2\Zab^4 m_R^2/M$.} \label{tab2}
\end{center}
\end{table}

It is interesting to compare results obtained with the Grotch-type
calculations in this paper with the Breit-type calculation
we perform previously \cite{a2Za4}.

A comparison of the Grotch-type results with the complete Breit-type
ones is summarized in Table~\ref{tab3}. The Breit-type results are
exact in $m/M$, while the Grotch-type recoil correction includes
only a term linear in $m/M$. As explained in \cite{II,a2Za4} (see
also \cite{pra}), one can rearrange the Breit Hamiltonian and
separate the linear recoil and higher-order terms. As stated
in \cite{a2Za4}, the linear recoil terms in the Breit-type approach
are consistent with the results obtained here; in fact they agree within
an uncertainty of numerical integration and, therefore, all digits
in the results given for the Grotch-type evaluation in Table~\ref{tab3}
are valid.

 \begin{table}
\begin{center}
\begin{tabular}{|c|l|l|l|l|}
\hline
Atom & {$~~1s$} & {$~~2s$} & {$~~~2p_{1/2}$}& $~~~2p_{3/2}$\\
\hline
H          & 0.172 & 0.0288 & 0.000488 & 0.000488 \\
           & {\em 0.155}&  {\em 0.0259} &  {\em 0.000737}&{\em 0.000289} \\
D          & 0.0973 & 0.0165 & 0.000290 & 0.000290 \\
           & {\em 0.0921}& {\em 0.0156} &  {\em 0.000370}&{\em 0.000227} \\
$^3$He$^+$ & 1.31 & 0.259 & 0.00621 & 0.00621 \\
           & {\em 1.26} & {\em 0.250} &  {\em 0.00810}&{\em 0.00492} \\
$^4$He$^+$ & 1.00 & 0.200 & 0.00478 & 0.00478 \\
           & {\em 0.97} & {\em 0.194} &  {\em 0.00590}&{\em 0.00403} \\
\hline
\end{tabular}
\caption{The $\alpha^2$ relativistic recoil eVP corrections to
energies of light muonic hydrogen-like atoms in units of $\mu$eV.
The eVP2 results of the Grotch-type evaluation of this paper are
given in roman type. The complete results of the Breit-type calculations
\cite{a2Za4} are presented in italics. Note that the complete Breit-type
recoil results \cite{a2Za4} of order $\alpha^2(Z\alpha)^4m$ are
exact in $m/M$, while the Grotch-type recoil contributions include
only terms linear in $m_R/M$.} \label{tab3}
\end{center}
\end{table}

As one can see from Table~\ref{tab3}, the higher-order terms  in $m/M$
are important for the complete recoil results. For the $s$
states in muonic hydrogen they are about 10\% of the linear term.
(Here in Table~\ref{tab3} the Darwin-Foldy-type terms are included for all
atoms. If, following \cite{spin0,spin1}, we exclude them, the results are
shifted and become 0.0635~$\mu$eV ($1s$), 0.0126~$\mu$eV ($2s$) for
muonic deuterium and 0.75~$\mu$eV ($1s$), 0.172~$\mu$eV ($2s$) for muonic
helium-4).
For the $p$ states the $(m/M)^2$ contribution is even larger in
fractional units, however, the total recoil contribution for the
$2p$ state is small in comparison with the related $2s$ contribution
and can be neglected for the $2p-2s$ difference, while calculating
the Lamb shift. A similar situation actually also takes place
for the one-loop eVP contribution \cite{pra,II}.

The Breit-type calculations \cite{a2Za4} delivered all the
contributions within a nonrelativistic perturbation theory (NRPT)
with various relativistic perturbations of the Coulomb potential.

Within the Breit-equation approach, the recoil and non-recoil terms
of the Breit Hamiltonian are treated in the same way (see, e.g.,
\cite{breit1}). As a result, the technique applied in \cite{a2Za4}
to obtain the relativistic non-recoil term (i.e. the relativistic
correction to the one-particle equation with the reduced mass) was
the same as for the recoil term. Actually, within the NRPT approach
there is no need for a separation between the recoil and non-recoil
terms.

Here, the recoil correction is obtained in a quite different
way. Thus, we conclude that our NRPT calculation of both the
non-recoil and recoil relativistic contributions \cite{a2Za4} is
correct.

To conclude, the results of this paper include the development of a
method to calculate the second-order-eVP relativistic recoil
correction for an arbitrary state in an arbitrary hydrogen-like
muonic atom. That is a purely theoretical result. As for an
application to practically important transitions, we have calculated
the relativistic recoil correction of order
$\alpha^2(Z\alpha)^4m^2/M$ for the Lamb shift in muonic hydrogen. It
is small by itself. As we explain above, such a calculation serves as
a confirmation of our eVP results previously obtained by means of
NRPT \cite{a2Za4}. Because of the way relativistic
recoil and non-recoil contributions were treated there, the result of this paper
confirms the whole relativistic eVP contribution of \cite{a2Za4} and
in particular its relativistic non-recoil correction of order
$\alpha^2(Z\alpha)^4m$. That correction, in contrast to the recoil
term, is somewhat smaller than, but still compatible with the
uncertainty of actual experiments (see \cite{a2Za4} for details).

\section*{Acknowledgments}

This work was supported in part by DFG under Grant No. GZ: HA 1457/7-2
and RFBR under Grant No. 12-02-31741. Part of the work was done
while V.G.I. and E.Y.K. stayed at  the Max-Planck-Institut f\"ur
Quantenoptik, and they are grateful for the warm hospitality they received.

\appendix

\section{Approximation for the irreducible part of the
K\"allen-Sabry dispersion function with an uncertainty better than 1\%\label{a:ks}}

The exact expression for the irreducible part of the dispersion
weight function of the K\"allen-Sabry potential
\cite{Schwinger,Kallen-Sabry,ro2hfs}
\begin{eqnarray}\label{def:rho2}
  \rho_{2}(v)&=&\frac{2}{3}\left(\frac{\alpha}{\pi}\right)^2\frac{v}{1-v^2}\times\biggl\{(3-v^2)(1+v^2)\nonumber\\
   &&\left[{\rm Li}_2\left(-\frac{1-v}{1+v}\right)+2{\rm Li}_2\left(\frac{1-v}{1+v}\right)\right.\nonumber\\
  &+&\left. \ln\left(\frac{1+v}{1-v}\right)\left(\frac{3}{2}\ln\left(\frac{1+v}{2}\right)  -\ln\left(v\right)\right)\right]\nonumber\\
  &+&\left(\frac{11}{16}(3-v^2)(1+v^2)+\frac{1}{4}v^4\right)\ln\left(\frac{1+v}{1-v}\right)\nonumber\\
  &+&\frac{3}{2}v(3-v^2)\ln\left(\frac{1-v^2}{4}\right)-2v(3-v^2)\ln(v)\nonumber\\
  &+&\frac{3}{8}v(5-3v^2)\biggr\}
\end{eqnarray}
is somewhat complicated. It does not allow us exact analytic evaluations in
muonic atoms. Meanwhile an approximate
representation by Schwinger \cite{Schwinger}
\begin{equation}
\rho^{\rm (s)}_2(v)=\left(\frac{\alpha}{\pi}\right)^2
\frac{v^2(1-\frac{v^2}{3})}{1-v^2} \left\{ \frac{\pi^2}{2v} -
\frac{3+v}{4} \left( \frac{\pi^2}{2}- \frac34 \right)
  \right\}
\end{equation}
is a well-known successful approximation. It allows us to
find contributions of the irreducible two-loop vacuum polarization
to various values with a high accuracy. Since it
reproduces the correct behavior of the dispersion density at $v\simeq0$
(i.e., for $s\simeq s_{\rm threashold}$) and $v\simeq
1$ (i.e.. for $s\to \infty$), it may be considered as an
extrapolation formula. Because of that it is useful not only to
approximately find various numeric contributions, but also to
approximate certain asymptotics.

Here, we present a few more extrapolations which in certain
respects are more accurate than Schwinger's \cite{Schwinger}.
They are
\begin{eqnarray}
\rho^{(1)}_2(v)&=&\left(\frac{\alpha}{\pi}\right)^2 \frac{v}{1-v^2}\nonumber\\
&\times&\biggl\{\frac{1}{2}v^2+(1-v^2)\left(\frac{\pi^2}{2}-2.62\,v\right)\biggr\}\;,
\\
\rho^{(2)}_2(v)&=&\left(\frac{\alpha}{\pi}\right)^2 \frac{v}{1-v^2}\nonumber\\
&&\times\biggl\{\left(\frac{1}{2}+0.288(1-v^2)\right)v^2\nonumber\\
&&+(1-v^3)\left(\frac{\pi^2}{2}-3.695\,v\right)\biggr\}\;,
\\
\rho^{(3)}_2(v)&=&\left(\frac{\alpha}{\pi}\right)^2 \frac{v}{1-v^2}\nonumber\\
&&\times\biggl\{\left(\frac{1}{2}+1.08(1-v^2)\right)v^2\nonumber\\
&&+(1-v^3)\left(\frac{\pi^2}{2}-3.97\,v-0.28\,v^3\right)\biggr\}\;.
\end{eqnarray}

The quality of these approximations can be discussed in the following
way. We note that the density function is always positive. If we are
to calculate a matrix element which does not change sign, such
as an average of the irreducible part of the K\"allen-Sabry
potential over a certain state, then the fractional error cannot
exceed the maximal fractional error of the approximation of
$\rho_{2}(v)$ in (\ref{def:rho2}) by an approximate function. In
general, if there are not specific cancelations, the fractional error
determines the fractional uncertainty of any integrals.

\begin{figure}
\begin{center}
\includegraphics[width=0.40\textwidth]{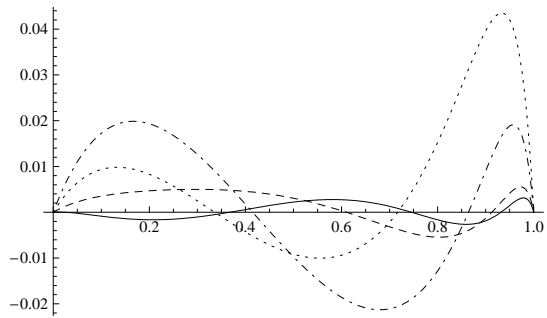}
\caption{Fractional deviation $\delta^{\rm approx}(v)$ of the
approximations from the exact dispersion density $\rho_2(v)$. The
dotted line presents a deviation for the original Schwinger extrapolation $\rho^{\rm (s)}_2$, the dot-dashed
line is for $\rho^{(1)}_2$, the dashed is for $\rho^
{(2)}_2$, and the
fractional deviation for $\rho^{(3)}_2$ is plotted as a solid line.} \label{fig:approx}
\end{center}
\end{figure}

Such an error
\[
\delta^{\rm approx}(v)=\frac{\rho^{\rm
approx}_2(v)-\rho_{2}(v)}{\rho_{2}(v)}
\]
is plotted for the considered approximations in
Fig.~\ref{fig:approx} along with the Schwinger approximation.
The maximal values $\delta^{\rm max}$  of the fractional deviation
are collected in Table~\ref{tab:delta}.

\begin{table}
\begin{center}
\begin{tabular}{|c|c|}
\hline
Approximation & $\delta^{\rm max}$\\
\hline
$\rho^{\rm (s)}_2(v)$     & 4\% \\
$\rho^{(1)}_2(v)$     & 2\% \\
$\rho^{(2)}_2(v)$      & 0.55\% \\
$\rho^{(3)}_2(v)$      & 0.3\% \\
\hline
\end{tabular}
\caption{The maximal values $\delta^{\rm max}$  of the fractional
deviation $\delta^{\rm approx}(v)$ for various approximations.} \label{tab:delta}
\end{center}
\end{table}

We note that the accuracy of polynomial approximations for the expression in the large parentheses in Eq. (A1) is limited

We note that the accuracy of polynomial approximation for the
expression in curly brackets in Eq.~(\ref{def:rho2}) is limited. It
is clear that the behavior close to $v=0$ and $v=1$ should include
logarithmic factors $\ln(v)$ and $\ln(1-v)$, respectively. Those
cannot be approximated with polynomials. However, approximations of
$\rho_{2}(v)$ with uncertainty below 1\% are possible.

The application of the approximate formulas to the calculation of the
relativistic and relativistic recoil corrections is summarized in
Tables~\ref{tab4} and \ref{tab5}. The fractional errors do not
exceed the values of $\delta^{\rm max}$ collected in
Table~\ref{tab:delta}.

\begin{table}
\begin{center}
\begin{tabular}{|c|c|c|c|c|}
\hline
$\rho_2(v)$ & $1s_{1/2}$ & $2s_{1/2}$ & $2p_{1/2}$& $2p_{3/2}$\\
\hline
 Axact $\rho_2(v)$&$-0.5856$&$-0.1026$&$-0.03110$&$-0.003754$\\
$\rho^{\rm (s)}_2(v)$     &$-0.5909$&$-0.1035$&$-0.03124$&$-0.003763$\\
$\rho^{(1)}_2(v)$     &$-0.5844$&$-0.1024$&$-0.03100$&$-0.003754$\\
$\rho^{(2)}_2(v)$     &$-0.5861$&$-0.1027$&$-0.03114$&$-0.003759$\\
$\rho^{(3)}_2(v)$     &$-0.5859$&$-0.1027$&$-0.03112$&$-0.003756$\\
\hline
\end{tabular}
\caption{The relativistic contribution to the energies for the irreducible
part of the K\"allen-Sabry potential in
muonic hydrogen. Units are $(\alpha/\pi)^2\Zab^4 m_R$.
The numerical results are for the solution of the Dirac equation
with the reduced mass.  \label{tab4}}
\end{center}
\end{table}

\begin{table}
\begin{center}
\begin{tabular}{|c|c|c|c|}
\hline
$\rho_2(v)$ & $1s$ & $2s$ & $2p$\\
\hline
 Exact $\rho_2(v)$     & $0.1390$  & $0.05762$  & $0.002879$\\
$\rho^{\rm (s)}_2(v)$  & $0.1425$  & $0.05837$  & $0.002895$\\
$\rho^{(1)}_2(v)$      & $0.1401$  & $0.05765$  & $0.002868$\\
$\rho^{(2)}_2(v)$      & $0.1392$  & $0.05767$  & $0.002881$\\
$\rho^{(3)}_2(v)$      & $0.1391$  & $0.05765$  & $0.002880$\\
\hline
\end{tabular}
\caption{The relativistic recoil contribution to the energies
$\Delta E^{(1)}_{2}$ for the irreducible part of the
K\"allen-Sabry potential in muonic hydrogen in units of $(\alpha/\pi)^2\Zab^4 m_R^2/M$. } \label{tab5}
\end{center}
\end{table}

The application of the approximations could be useful not only for tests
but also for approximate analytic expressions (cf. \cite{uehl_cl,rel1,rel2,uehl_an}).

\section{Representation of the reduced nonrelativistic Coulomb Green function in coordinate space\label{a:g}}

The nonrelativistic Coulomb Green's function
\[
G_{E}({\bf r},{\bf r}^\prime)=\sum\frac{|\lambda \rangle\langle \lambda |}{E-E_\lambda}
\]
and its reduced form
\[
G^\prime_{E,nlm}({\bf r},{\bf r}^\prime)={\sum}'\frac{|\lambda \rangle\langle \lambda |}{E-E_\lambda}\;,
\]
where one has to sum over all intermediate states $\lambda$ for
$G_{E}$ and for all but the reference state $nlm$ for the
reduced one, have a number of useful representations.

It is helpful to separate the radial and angular parts for the partial
contributions in the full Green's function
\begin{eqnarray}\label{b1}
G_{E}({\bf r},{\bf r}^\prime)&=&\sum_{lm}\,{G}_{E,nl}(r,r^\prime)\,Y_{lm}^\star(\Omega)Y_{lm}(\Omega^\prime)\nonumber\\
&=& \frac{2l+1}{4\pi}\,{G}_{E, nl}(r,r^\prime)\,P_l(\cos\theta)\,,
 \end{eqnarray}
where $Y_{lm}(\Omega)$ are spherical harmonics, $m$ is the
projection of orbital momentum, $\Omega$ is the angular variable,
$P_l(\cos\theta)$ is the Legendre polynomial and $\theta$ is the angle
between ${\bf r}$ and ${\bf r}^\prime$. A similar separation can be
also done for the reduced Green functions.

In our calculations we deal with matrix elements as
\[
\langle nlm | A G^\prime_{E_{nlm}, nlm} B | nlm \rangle\;,
\]
where $A$ and $B$ are for central potentials. In such a case the
only partial contribution surviving in the sum over $lm$ in (\ref{b1})
is ${G}_{E,nlm}$. The result for the matrix element does not depend on $m$.
For further consideration we denote it as ${G}_{E,nl}$.
For the reduced Green function with $E=E_{nl}$, we denote the
surviving term in the partial sum as $G^\prime_{nl}$.

In our paper we apply two representations, which are described below.

\subsection{The Hostler presentation}

One of the efficient representations of the nonrelativistic Coulomb
Green function was derived by Hostler \cite{Hostler-1964}
 \begin{eqnarray}
 G_{E,nl}({\bf r},{\bf r}^\prime)&=&\frac{4Z\alpha\,m_r^2}{\nu\,z_>z_<}
 \frac{\Gamma(1+l-\nu)}{\Gamma(2l+2)}\,\nonumber\\
 &\times&W_{\nu,l+1/2}(z_>)M_{\nu,l+1/2}(z_<)\,,
 \end{eqnarray}
where
 \begin{eqnarray}
 \nu&=&\frac{Z\alpha\,m_r}{\sqrt{-2m_r E}}\,,\\
 z_>&=&\frac{2Z\alpha\,m_r}{\nu}{\rm max}(r,r^\prime)\,,\\
 z_<&=&\frac{2Z\alpha\,m_r}{\nu}{\rm min}(r,r^\prime)\,,
 \end{eqnarray}
$\Gamma(x)$ is the Gamma function, and $M_{\mu\nu}(x)$ and $W_{\mu\nu}(x)$
are the Whittaker functions
\cite{bookMW}.

The required radial parts of the reduced Coulomb Green's functions
of the states of interest are
\cite{hameka-1967,Hostler-1969,pachucki1996}
\begin{eqnarray}
{G}'_{1s}(r,r^\prime)&=&4Z\alpha\,m_r^2\,\exp{\left(-\frac{z_>+z_<}{2}\right)}\biggl\{\frac{1}{z_>}+\frac{1}{z_<}\nonumber \\
&~&+\frac{7}{2}-\frac{z_>+z_<}{2}+{\rm Ei}(z_<)-2C \nonumber \\
 &~&-\ln(z_>z_<)-\frac{e^{z_<}}{z_<}\biggr\}\,,
\end{eqnarray}
\begin{eqnarray}
{G}'_{2s}(r,r^\prime)&=&Z\alpha\,m_r^2\frac{\exp\left(-\frac{z_>+z_<}{2}\right)}{4z_>z_<}\biggl\{8z_<-4z_<^2+8z_>
\nonumber\\ \nonumber
 &+& 12 z_>z_< - 26z_>z_<^2 + 2z_>z_<^3 - 4z_>^2 \\ \nonumber
 &-& 26z_>^2z_< + 23z_>^2z_<^2 -z_>^2z_<^3 + 2z_>^3z_< \\ \nonumber
 &-& z_>^3z_<^2 + 4  (z_> - 2)z_>(1 - z_<)e^{z_<}  \\ \nonumber
 &+& 4 (z_> - 2)z_>(z_< - 2)z_< \\
 &\times& \bigl[-2C +{\rm Ei}(z_<) - \ln(z_>z_<)\bigr] \biggr\}\,,
 \end{eqnarray}
\begin{eqnarray}
{G}'_{2p}(r,r^\prime)&=&Z\alpha\,m_r^2\frac{\exp\left(-\frac{z_>+z_<}{2}\right)}{36(z_>z_<)^2}\biggl\{24z_<^3
+ 36z_>z_<^3  \nonumber\\ \nonumber
 &+& 36z_>^2z_<^3+ 24 z_>^3 + 36z_>^3z_< + 36z_>^3 z_<^2\\ \nonumber
 &+& 49  z_>^3z_<^3 - 3  z_>^3z_<^4-3z_>^4z_<^3  \\ \nonumber
 &-& 12 z_>^3(2+z_< +z_<^2) e^{z_<}+12z_>^3z_<^3 \\
 &\times&  \bigl[-2C +{\rm Ei}(z_<) - \ln(z_>z_<)\bigr]\biggr\}\,,
\end{eqnarray}
where $C=0.577\,216\ldots$ is the Euler constant, and
\[
{\rm Ei}(x)=\int\limits_{-\infty}^x\frac{e^t}{t}dt
\]
is the exponential integral.

\subsection{The Sturmian representation}

The Sturmian presentation of the nonrelativistic Coulomb Green's
function is a presentation in terms of a basis set which consists of
solutions of the related Sturm-Liouville problem.

The basis functions satisfy the equation%
\begin{equation}\label{sturm}
{\bf p}^2\Phi_{klm}(\nu;\;{\bf
r})=2m_r\left(\frac{k}{\nu}\,\frac{Z\alpha}{r}+E\right)\Phi_{klm}(\nu;\;{\bf
r})\,,
\end{equation}
where
\[
  \nu=\sqrt{-\frac{(\Za)^2m}{2E}}
  \,.
\]

They can be presented in the form
\[
\Phi_{klm}(\nu;\;{\bf r})=R_{kl}(\nu;r)Y_{lm}(\Omega)\,,
\]
where
\[
 R_{kl}(\nu;\;r)=\left(\frac{k}{\nu}\right)^{3/2}R_{kl}\left(\frac{k}{\nu}\,r\right)\;,
\]
and $R_{kl}(r)$ stands for the radial part of the standard wave function
of the nonrelativistic Coulomb problem (see, e.g., \cite{III}).

The radial part of the Coulomb Green's function is of the form
\cite{sturm}
\begin{equation}
  G_{El}(r;r')  =
  \frac{\nu^2}{(Z\alpha)^2m}\sum_{k=l+1}^{\infty}\frac{k}{k-\nu}
  R_{kl}\left(\nu;\;r\right)  R_{kl}\left(\nu;\;r'\right) \,.
\end{equation}

The required radial part of the reduced Coulomb Green's function (at $E=E_n$) is of
the form \cite{sturm}
 \begin{eqnarray}
 {G}^\prime_{nl}(r,r^\prime)&=&\frac{n^2}{(Z\alpha)^2m_r}
 \Biggl\{ \sum_{k=l+1\atop k\neq n}^\infty \frac{k}{k-n}R_{kl}(n;r)R_{kl}(n;r^\prime)
 \nonumber\\
 &+& \frac{3}{2}R_{nl}(n;r)R_{nl}(n;r^\prime) \nonumber\\
 &+&  rR^\prime_{nl}(n;r)R_{nl}(n;r^\prime)\nonumber\\
 &+&r^\prime R^\prime_{nl}(n;r^\prime)R_{nl}(n;r)  \Biggr\} \,,\label{gred}
  \end{eqnarray}
where
\[
R^\prime_{nl}(n;r)=\frac{\partial}{\partial r}R_{nl}(n;r)\;.
\]

\end{document}